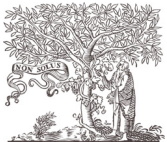
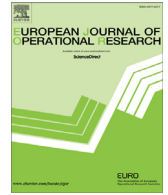

Invited Review

# Estimating causal effects with optimization-based methods: A review and empirical comparison

Martin Cousineau[a], Vedat Verter[b,*], Susan A. Murphy[c], Joelle Pineau[d]

[a] *HEC Montréal, Department of Logistics and Operations Management, 3000, chemin de la Côte-Sainte-Catherine, Montréal, Québec H3T 2A7, Canada*
[b] *Michigan State University, Broad College of Business, Department of Supply Chain Management, 632 Bogue St., East Lansing, Michigan 48824, USA*
[c] *Harvard University, John A. Paulson School of Engineering and Applied Sciences, 29 Oxford St., Cambridge, Massachusetts 02138, USA*
[d] *McGill University, School of Computer Science, 3480 University St., Montréal, Québec H3A 0E9, Canada*

## ABSTRACT

In the absence of randomized controlled and natural experiments, it is necessary to balance the distributions of (observable) covariates of the treated and control groups in order to obtain an unbiased estimate of a causal effect of interest; otherwise, a different effect size may be estimated, and incorrect recommendations may be given. To achieve this balance, there exist a wide variety of methods. In particular, several methods based on optimization models have been recently proposed in the causal inference literature. While these optimization-based methods empirically showed an improvement over a limited number of other causal inference methods in their relative ability to balance the distributions of covariates and to estimate causal effects, they have not been thoroughly compared to each other and to other noteworthy causal inference methods. In addition, we believe that there exist several unaddressed opportunities that operational researchers could contribute with their advanced knowledge of optimization, for the benefits of the applied researchers that use causal inference tools. In this review paper, we present an overview of the causal inference literature and describe in more detail the optimization-based causal inference methods, provide a comparative analysis of the prevailing optimization-based methods, and discuss opportunities for new methods.

## 1. Introduction

*Randomized experiments* are the gold standard for ascertaining the efficacy of an intervention. In such experiments, the participants are randomly assigned into an experimental group (aka the treated group) or a control group. This guarantees that the expected difference in outcomes would be due only to the difference in treatment. It also ensures that the treated and control groups are similar in the distribution of their (observable as well as unobservable) covariates. This similarity is necessary in order to make an unbiased comparison of these two groups. Otherwise, a different treatment effect may be estimated, leading to incorrect recommendations concerning the intervention. Unfortunately, it is not always practical or deemed ethical to run such randomized controlled experiments; these are known to have several limitations such as their cost and external validity (Black, 1996; Sanson-Fisher, Bonevski, Green, & D'Este, 2007). In these cases, using observational data is the only viable option that needs to be exercised carefully. Since the distribution of the covariates may differ between the treated and control groups in observational data, it is necessary to adjust for this potential bias using study designs or methods from the field of causal inference.

A noteworthy alternative to randomized controlled experiments which use observational data consists of *natural experiments*. These include (1) studies in which the treatment assignment is randomized not for the purpose of a study (e.g., a school lottery such as in Cullen, Jacob, & Levitt, 2006), (2) studies with an instrumental variable, i.e., a randomized variable upstream of the treatment assignment (e.g., KC & Terwiesch, 2012), (3) regression discontinuity designs (e.g., Calvo, Cui, & Serpa, 2019), and (4) difference-in-difference designs (e.g., Gallino & Moreno, 2014). Unfortunately, similarly to randomized controlled experiments, these natural experiments may also be infeasible since the causal variable may not be randomized, a strong instrumental variable may not be available, the treatment assignment may not be based on a measured cutoff value, or no panel data may be available, respectively.

* Corresponding author.
  *E-mail addresses:* martin.cousineau@hec.ca (M. Cousineau), verter@broad.msu.edu (V. Verter), samurphy@fas.harvard.edu (S.A. Murphy), joelle.pineau@mcgill.ca (J. Pineau).

*Causal inference* methods constitute the next viable option to eliminate potential bias from observational data when the two avenues mentioned above are not feasible. Traditionally, these techniques are typically grouped as *stratification*, *matching*, *weighting* and *regression*. Note however that in order to remove all potential bias in an observational study, these methods need to assume, among other things, that they are using adequate covariates, i.e., covariates with which the ignorability assumption holds (see Section 2); while it is possible to perform sensitivity analyses of this claim, it is not possible to test it without further information. There exist a wide variety of such methods in the literature, scattered across many disciplines such as statistics (Rosenbaum, 2002; Rubin, 2006), economics (Imbens, 2004), epidemiology (Brookhart et al., 2006), sociology (Morgan & Harding, 2006), political science (Ho, Imai, King, & Stuart, 2007), artificial intelligence (Pearl, 2009b; Sun & Nikolaev, 2016), and notably operational research (Kwon, Sauppe, & Jacobson, 2019; Nikolaev, Jacobson, Cho, Sauppe, & Sewell, 2013; Sauppe, Jacobson, & Sewell, 2014).

Recently, several *optimization-based* causal inference methods have been proposed, which try to identify weights that, when applied to the subjects, remove the bias from the estimated effect. A first advantage of such methods is that they can guarantee that the returned solution optimizes a criterion (i.e., the objective function) and complies with a number of conditions (i.e., the constraints). In other words, these methods can guarantee that the identified weights are the best to estimate the causal effect, if one agrees a priori with the objective function and the constraints. In addition, these optimization-based methods generally allow to separate the "design" and "analysis" stages of these observational studies. Similar to an experimental study, only background information on the subjects (i.e., the covariates) should be used when designing the study (i.e., when eliminating potential bias), and the outcomes should only be used during the analysis stage (i.e., when estimating the treatment effect) (Rubin, 2007; Stuart, 2010).

We believe that there still exist several opportunities for the development of new causal inference methods based on optimization (e.g., mathematical programming), a technique well-known to operational researchers. Thus, the main goal of this review paper is to encourage further contributions from the operational research community to the causal inference literature. First, for the lay researchers, we provide an overview of the scattered causal inference literature. Then, we discuss in more detail the existing optimization-based causal inference methods and provide an empirical comparison of these methods. To our knowledge, such an extensive comparison of these optimization-based methods is not currently available in the literature. Finally, to guide future research, we discuss relevant opportunities for the proposition of new optimization-based methods.

For the sake of clarity and space, the scope of this non-systematic review and empirical comparison is mainly focused on methods familiar to operational researchers (e.g., mathematical programming, heuristics). Hence, we do not cover in detail all of the different streams of research on causal inference; for example, we do not cover the methods based on neural networks (e.g., Johansson, Shalit, & Sontag, 2016; Shalit, Johansson, & Sontag, 2017; Louizos et al., 2017; Yao et al., 2018; Kallus, 2020) even if the training of such neural networks requires optimization. We do, however, try to acknowledge some of these methods where relevant to show the different possibilities. In order to identify the relevant studies to discuss, a search was performed within the full text for "causal inference", "causal effect" and "treatment effect" with Google Scholar. Then, the titles and abstracts were reviewed to identify relevant studies to include, with an emphasis on peer-reviewed journal articles and conference proceedings. Additional studies were also included by backward snowballing. Once the optimization-based methods were identified, a search was then performed to find the public implementations of these methods (if available).

The remainder of this paper is organized as follows. Section 2 introduces the notation and assumptions. Section 3 motivates the need for causal inference methods with an illustrative example. Section 4 provides a selective overview of causal inference methods. Section 5 describes the prevailing optimization-based methods. Section 6 provides comparative results of several causal inference methods, including several optimization-based methods. Section 7 discusses research opportunities. Section 8 concludes the paper with final remarks.

## 2. Foundations

Following Neyman-Rubin's potential outcome framework (Rubin, 2005), let $Z_i$ denote a binary treatment, with $Z_i = 0$ indicating the assignment of subject $i$ to control and $Z_i = 1$ indicating the assignment of subject $i$ to treatment. Let $Y_i(0)$ and $Y_i(1)$ denote the potential outcomes, i.e., the outcomes that would occur for subject $i$ if $Z_i = 0$ and $Z_i = 1$, respectively. Let $Y_i$ denote the *observed outcome* for subject $i$; $Y_i = (1 - Z_i)Y_i(0) + Z_iY_i(1)$. Let $X_i$ denote a vector of pre-treatment covariates for subject $i$. We also define the following sets:

- $\mathcal{I}$: the sample of $n$ subjects,
- $\mathcal{I}_{Z=0} = \{i : Z_i = 0\}$: the set of control subjects
- $\mathcal{I}_{Z=1} = \{i : Z_i = 1\}$: the set of treated subjects, and
- $\mathcal{I}_{X=x} = \{i : X_i = x\}$: the set of subjects having covariates $x$.

Using this notation, several common estimands can be defined where the choice of the estimand of interest depends on the aim of the study and data availability. We now describe estimators for three such estimands. First, the estimator for the sample average treatment effect (SATE),

$$\tau_{SATE} = \frac{1}{|\mathcal{I}|} \sum_{i \in \mathcal{I}} [Y_i(1) - Y_i(0)] \tag{1}$$

is used to compute the effect of a treatment on a sample, with $|\cdot|$ denoting the cardinality of a set. Then, the estimator for the conditional average treatment effect (CATE),

$$\tau_{CATE}(x) = \frac{1}{|\mathcal{I}_{X=x}|} \sum_{i \in \mathcal{I}_{X=x}} [Y_i(1) - Y_i(0)] \tag{2}$$

is used to compute the effect of a treatment on a subgroup defined by some covariates $x$. Finally, the estimator for the sample average treatment effect on the treated (SATT),

$$\tau_{SATT} = \frac{1}{|\mathcal{I}_{Z=1}|} \sum_{i \in \mathcal{I}_{Z=1}} [Y_i(1) - Y_i(0)] \tag{3}$$

is used to compute the effect of a treatment on the subgroup that received the treatment. SATT makes sense in a setting such as a social program in which only a subpopulation may come forward to benefit from this program.

It is not possible to observe both potential outcomes $Y_i(0)$ and $Y_i(1)$ for a subject $i$, because each subject is either in the control or the treated group. Therefore, some assumptions are required in order to estimate the treatment effects mentioned above. Two typical assumptions are *ignorability* and *overlap*, which are often jointly referred to as *strong ignorability* (Rosenbaum & Rubin, 1983). Note that the ignorability assumption is known under "selection on observables", "all confounders measured", "exchangeability", "conditional independence assumption", "no hidden bias" and "unconfoundness", whereas the overlap assumption is also known as "common support" and "probabilistic assignment assumption" (Dorie, Hill, Shalit, Scott, & Cervone, 2019; Li, Morgan, & Zaslavsky, 2018).

**Assumption 1** (Strong ignorability). Conditional on the covariates, the treatment assignment is independent of the potential outcomes. In addition, conditional on the covariates, the treatment assignment is always possible. Formally, for all $X$ in the sample, assume

$$Y(0), Y(1) \perp\!\!\!\perp Z \mid X \quad \text{(ignorability)} \tag{4}$$

and

$$0 < \Pr(Z = 1 \mid X) < 1 \quad \text{(overlap)} \tag{5}$$

holds, with $\perp\!\!\!\perp$ denoting independence. It is also possible to state these assumptions using the language of causal graphs (Pearl, 2009a).

Under the strong ignorability assumption, it is then possible to estimate the previous treatment effects after, for example, obtaining similar distributions of covariates for the control and treated groups. We refer to this process as balancing the treated and control groups.

In the remainder of this paper, we assume that strong ignorability holds. Note, however, that this is a strong assumption, which may not be satisfied in practice. In particular, in order for ignorability to hold, applied researchers need to identify the relevant covariates with the use of domain knowledge and their understanding of the causal relationships between the variables. If they incorrectly identify these covariates, do not have such covariates at hand or only rely on a convenient set of covariates, then the methods may show poor performance (e.g., Shadish, Clark, & Steiner, 2008). Finally, it is not possible to test this assumption from observational data without further information. This is a major limitation of the prevailing causal inference methods, sometimes alleviated through the use of sensitivity analyses (Li, 2013; Stuart, 2010). Still, this assumption (or a variant of it) is necessary when comparing alternative methods (e.g., Dorie et al., 2019; Hahn, Dorie, & Murray, 2019).

## 3. An illustrative example

To highlight the need for causal inference, we now provide an illustrative example. We have a population of patients in which we want to compute the effect of a treatment. For the purposes of this example, suppose half of the patients have high social support ($X = 1$). These patients have a 60% response rate and patients with low social support ($X = 0$) have a 40% response rate. Note that a patient's response or outcome is independent of whether she is treated ($Z = 1$) or not ($Z = 0$), i.e., in reality the treatment has no effect. Thus, $E[Y \mid Z = z, X = 1] = 0.6$ and $E[Y \mid Z = z, X = 0] = 0.4$ for $z = 0, 1$.

When we fully randomize the treatment to the patients, we obtain a treated and a control group that contain equal proportions of patients with low and high social support, respectively. We thus obtain the following response rates for the controls

$$E[Y \mid Z = 0] = \sum_{x=0}^{1} E[Y \mid Z = 0, X = x] \Pr(X = x \mid Z = 0)$$
$$= (.4)(.5) + (.6)(.5)$$
$$= 0.5$$

and for the treated

$$E[Y \mid Z = 1] = \sum_{x=0}^{1} E[Y \mid Z = 1, X = x] \Pr(X = x \mid Z = 1)$$
$$= (.4)(.5) + (.6)(.5)$$

**Table 1**
Physician's propensity to treat given social support covariate.

| Treatment | Low social support $X = 0$ | High social support $X = 1$ |
|---|---|---|
| $Z = 0$ | $\Pr(Z = 0 \mid X = 0) = 0.5$ | $\Pr(Z = 0 \mid X = 1) = 0.1$ |
| $Z = 1$ | $\Pr(Z = 1 \mid X = 0) = 0.5$ | $\Pr(Z = 1 \mid X = 1) = 0.9$ |

$$= 0.5.$$

In a randomized controlled experiment, the data provide an unbiased estimate of the true treatment effect - in this case, a null effect. We obtain this result since the treated and control groups are balanced by design, i.e., the randomization ensures that (large enough) treated and control groups have similar distributions of covariates.

Lets now suppose that it is not possible to run a randomized controlled experiment and instead we are given observational data by a clinician. Suppose that this clinician *incorrectly* suspects that patients with high social support ($X = 1$) benefit more from the treatment. Thus, this clinician treats 90% of his patients with high social support ($\Pr(Z = 1 \mid X = 1) = 0.9$) and only 50% of his patients with low social support ($\Pr(Z = 1 \mid X = 0) = 0.5$). Table 1 depicts the resulting parameters.

If we use the observational data, carelessly, to estimate the response rates, we obtain the following for the controls

$$E[Y \mid Z = 0] = \frac{\sum_{x=0}^{1} E[Y \mid Z=0, X=x] \Pr(Z=0 \mid X=x) \Pr(X=x)}{\sum_{x=0}^{1} \Pr(Z=0 \mid X=x) \Pr(X=x)}$$
$$= \frac{(.4)(.5)(.5) + (.6)(.1)(.5)}{(.5)(.5) + (.1)(.5)}$$
$$= 0.43$$

and for the treated

$$E[Y \mid Z = 1] = \frac{\sum_{x=0}^{1} E[Y \mid Z=1, X=x] \Pr(Z=1 \mid X=x) \Pr(X=x)}{\sum_{x=0}^{1} \Pr(Z=1 \mid X=x) \Pr(X=x)}$$
$$= \frac{(.4)(.5)(.5) + (.6)(.9)(.5)}{(.5)(.5) + (.9)(.5)}$$
$$= 0.53.$$

We easily see that we do not obtain the true treatment effect, since the treated and control groups are not balanced anymore with respect to the social support covariate. We obtain a treatment effect of 0.1 while the true effect is null.

If we instead *balance the treated and control groups* with respect to $X$, then in the balanced groups the treatment would be independent of the social support, i.e., $\Pr(Z = 0 \mid X = x) = \Pr(Z = 0)$ and $\Pr(Z = 1 \mid X = x) = \Pr(Z = 1)$. This balancing can be done by modifying the weight of the different patients in each group such that the new respective distributions of the covariates are similar, i.e., $\Pr(X = x \mid Z = 0) = \Pr(X = x \mid Z = 1)$. The details of how to balance these groups will be the subject of the next sections. We thus obtain the following corrected response rates for the controls

$$E[Y \mid Z = 0] = \frac{\sum_{x=0}^{1} E[Y \mid Z=0, X=x] \Pr(Z=0 \mid X=x) \Pr(X=x)}{\sum_{x=0}^{1} \Pr(Z=0 \mid X=x) \Pr(X=x)}$$
$$= \frac{\Pr(Z=0)[(.4)(.5) + (.6)(.5)]}{\Pr(Z=0)[(.5) + (.5)]}$$
$$= 0.5$$

and for the treated

$$E[Y \mid Z = 1] = \frac{\sum_{x=0}^{1} E[Y \mid Z=1, X=x] \Pr(Z=1 \mid X=x) \Pr(X=x)}{\sum_{x=0}^{1} \Pr(Z=1 \mid X=x) \Pr(X=x)}$$
$$= \frac{\Pr(Z=1)[(.4)(.5) + (.6)(.5)]}{\Pr(Z=1)[(.5) + (.5)]}$$

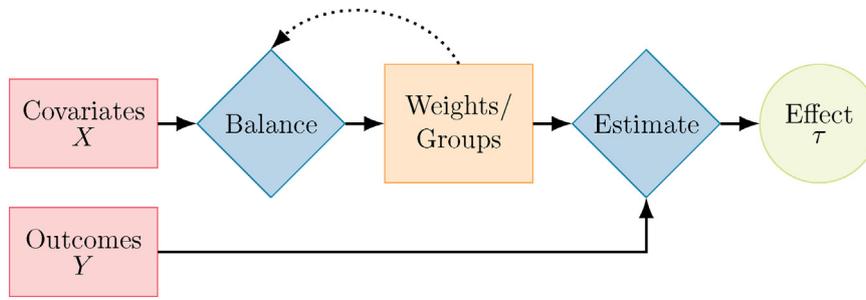

**Fig. 1.** Balancing methods.

$= 0.5$.

This indeed corresponds to the true treatment effect. Causal inference methods enable us to achieve the required balance when computing a treatment effect from observational data.

Using the potential outcome notation of Section 2 on the observational data, we have that $Y(0), Y(1)$ are independent of $Z$ conditional on $X$. It is easy to see here because the outcome depends only on $X$. For example, $E[Y(0) \mid Z = 0, X = 0] = E[Y(0) \mid X = 0] = 0.4$. However, $Y(0), Y(1)$ are not marginally independent of $Z$. For example, $E[Y(0) \mid Z = 0] = E[Y \mid Z = 0] = 0.43$ while $E[Y(0)] = 0.5 * 0.4 + 0.5 * 0.6 = 0.5$. After balancing the treated and control groups with respect to $X$, $Y(0), Y(1)$ are marginally independent of $Z$. In the balanced data, we have $E[Y(0) \mid Z = 0] = E[Y \mid Z = 0] = 0.5$ which is the same as $E[Y(0)] = 0.5$.

## 4. Overview of causal inference methods

In this section, we review a variety of methods that can be used for causal inference. In our selective review, we categorize the methods as balancing methods, regression methods and mixed methods. We will refer to this taxonomy in the next section when discussing the optimization-based methods.

### 4.1. Balancing methods

The balancing methods are useful to preprocess data in order to balance the distribution of the covariates between the control and treated subjects. As illustrated in Fig. 1, balancing methods only use the covariates $X$ to obtain balancing weights (e.g., propensity scores) or balanced groups (e.g., matched groups), and then estimate the treatment effect $\tau$ with the use of the outcomes $Y$. When used as a preprocessing step to a regression method, these methods can reduce the model dependence of the treatment effect estimate (Ho et al., 2007). In fact, if sufficiently balanced groups are obtained, a simple (weighted) difference in means may suffice to estimate these treatment effects. In addition, since balancing methods do not use the observed outcomes when preprocessing the data, they allow to test different parameter values until sufficiently balanced groups are obtained (see the dotted arrow in Fig. 1), while still ensuring the objectivity of the result (Rubin, 2007). The balancing methods, as defined here, include stratification, matching and weighting. Note that our balancing methods' definition assumes that the estimation of the effect is done through a simple model such as a (weighted) difference in means; more complex estimation procedures are part of the regression and mixed methods.

Several of these methods often use the propensity score (i.e., the probability of treatment assignment, $\pi_i = \Pr(Z_i = 1 \mid X_i)$) in order to do this balancing, since the propensity score is known to be a balancing score (Rosenbaum & Rubin, 1983). In particular, the propensity score is known to be the coarsest balancing score while the covariates $X$ are known to be the finest balancing score (Rosenbaum & Rubin, 1983). Yet, in the context of an observational study, the propensity score is not readily available and must be estimated from data. In addition to logistic regression, several methods have been proposed, such as tree-based methods including their bagged and boosted variants (Lee, Lessler, & Stuart, 2010; McCaffrey, Ridgeway, & Morral, 2004; Setoguchi, Schneeweiss, Brookhart, Glynn, & Cook, 2008), random forests (Lee et al., 2010), and neural networks (Setoguchi et al., 2008).

*Stratification* methods divide the treated and control subjects into a small number of subgroups. By doing so, it is possible to compare the observed outcomes of the treated and control subjects within a particular subgroup (Cochran, 1968). Yet, doing stratification on the covariates $X$ can become problematic as the number of covariates increases (Cochran, 1965). To address this issue, methods that stratify using the propensity score have also been proposed (Rosenbaum & Rubin, 1984). With stratification, the effects are generally estimated within each subgroup and then aggregated across subgroups to estimate the SATE or SATT.

*Matching* methods use the most similar subjects in the control group to infer the unobserved potential outcome of a subject in the treated group, and vice versa. Different variants of matching exist around the definition of distance (i.e., the similarity measure), the numbers of subjects in a group that can be matched to a subject in another group and whether subjects can be matched multiple times (Stuart, 2010). Some examples of these variants are Mahalanobis matching (Cochran & Rubin, 1973; Rubin, 1979; 1980), propensity score matching (Rosenbaum & Rubin, 1983), Mahalanobis matching within propensity score calipers (Rubin & Thomas, 2000), full matching (Hansen, 2004; Rosenbaum, 1991; Stuart & Green, 2008), fine balancing (Rosenbaum, Ross, & Silber, 2007), coarsened exact matching (Iacus, King, & Porro, 2012), genetic matching (Diamond & Sekhon, 2013; Sekhon & Grieve, 2007), almost-exact matching (Dieng, Liu, Roy, Rudin, & Volfovsky, 2019; Parikh, Rudin, & Volfovsky, 2021; Wang et al., 2021) and adaptive hyper-box matching (Morucci, Orlandi, Roy, Rudin, & Volfovsky, 2020). Note that several of these variants typically use a network algorithm to match subjects (e.g., see the R package optmatch). In general, these matching methods use the aggregated matched data and not the matched pairs to estimate the SATT; they can also be used, however, to estimate the SATE.

*Weighting* methods balance the control and treated groups by applying weights to the control or treated subjects. These weights are generally obtained through the propensity score. For example, a basic variant of inverse probability of treatment weighting (Rosenbaum, 1987) uses the weights $w_{SATE,i} = \frac{Z_i}{\hat{\pi}_i} - \frac{1-Z_i}{1-\hat{\pi}_i}$, where $\hat{\pi}_i$ is the fitted propensity score, to estimate the SATE as in Imbens (2004):

$$\hat{\tau}_{SATE} = \frac{1}{|\mathcal{I}|} \sum_{i \in \mathcal{I}} w_{SATE,i} Y_i. \quad (6)$$

To estimate the SATT, it is possible to weight instead by the odds, e.g., $w_{SATT,i} = Z_i - \frac{(1-Z_i)\hat{\pi}_i}{1-\hat{\pi}_i}$ (Imbens, 2004). There exist several other weighting variants to estimate a treatment effect of interest. For example, since propensity score models can lead to extreme weights, methods using stabilized weights (Cole & Hernan, 2008), trimmed weights (Lee, Lessler, & Stuart, 2011) and overlap weights (Li et al., 2018) have been proposed to alleviate this issue. Finally, these weighting methods (and other balancing methods) can also be combined with regression methods, what we call mixed methods. We describe regression methods in the next section, followed by the mixed methods.

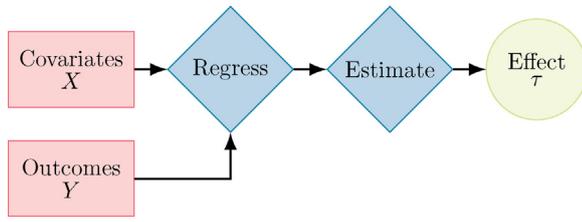

Fig. 2. Regression methods.

Table 2
Definition of the samples of subjects to estimate the weight vector(s).

| Causal effect | $w^{\text{treated}}$ | | $w^{\text{control}}$ | |
|---|---|---|---|---|
| | $\mathcal{U}$ | $\mathcal{V}$ | $\mathcal{U}$ | $\mathcal{V}$ |
| SATE | $\mathcal{I}_{Z=1}$ | $\mathcal{I}$ | $\mathcal{I}_{Z=0}$ | $\mathcal{I}$ |
| CATE(x) | $\mathcal{I}_{Z=1}$ | $\mathcal{I}_{X=x}$ | $\mathcal{I}_{Z=0}$ | $\mathcal{I}_{X=x}$ |
| SATT | | | $\mathcal{I}_{Z=0}$ | $\mathcal{I}_{Z=1}$ |

### 4.2. Regression methods

*Regression* methods directly model the response surface, i.e., they provide a model of the outcome over the subjects. As illustrated in Fig. 2, these methods both use the covariates $X$ and the outcomes $Y$ to model the response surface. Then, using this response surface, they can estimate the treatment effect $\tau$. Hence, in contrast with balancing methods, they do not clearly separate the "design" and "analysis" stages since they address the bias through their regression model (Stuart, 2010). In addition to this objectivity issue of the results, regression methods can be problematic if the model of the response surface is not flexible enough. Yet, if these methods are flexible enough, there is no need for preprocessing the data with balancing methods (Hill, 2011). Some examples of flexible regression methods include regression trees (Athey & Imbens, 2016; Hill, 2011; Taddy, Gardner, Chen, & Draper, 2016), random forests (Taddy et al., 2016; Wager & Athey, 2018), Bayesian adaptive LASSO (Ratkovic & Tingley, 2018) and deep neural networks (Johansson et al., 2016; Louizos et al., 2017; Shalit et al., 2017; Yao et al., 2018). Note that these flexible methods generally allow the estimation of the SATE, SATT and CATE.

### 4.3. Mixed methods

*Mixed* methods use both a balancing method and a regression method in order to estimate a treatment effect of interest; here, we consider a fitted model of the propensity score as a balancing method. Several of these methods can be represented by Fig. 3. where (1) the covariates $X$ are used to obtain some weights or groups, (2) these weights or groups are then used to model the response surface with the outcomes $Y$, and (3) the treatment effect $\tau$ is estimated. An early example of these methods is regression with propensity adjustment (Rosenbaum & Rubin, 1983) and doubly robust estimators (Bang & Robins, 2005). There also exist several other recent examples such as targeted maximum likelihood estimation (van der Laan & Rubin, 2006), approximate residual balancing (Athey, Imbens, & Wager, 2018) and augmented minimax linear estimation (Hirshberg & Wager, 2019).

## 5. Optimization-based methods

In this section, we describe in more detail the available optimization-based causal inference methods, most of which use mathematical programming. We also show the relationships between these model formulations and discuss the link that exists between these methods and other popular methods from the causal inference literature. Note that this section focuses on the mathematical aspects of these methods; the practical aspects of their implementation will be discussed in the next section along with their computational performance.

As previously discussed, these optimization-based methods are of interest since they allow to directly specify and optimize the characteristics pursued. This is in contrast with balancing methods that indirectly address balance and that may require several iterations, in which, for example, the propensity score model of a weighting method or the distance function of a matching method is adjusted. In addition, most of these optimization-based methods allow balancing without relying on the observed outcomes and therefore ensure the objectivity of the results in contrast to regression methods. Yet, an important potential issue of some of these optimization-based methods is that they could require long computational time to balance these covariates, in particular when having a large number of subjects. To reduce these computational times, some of these optimization-based methods rely on customized algorithms or on commercial solvers.

For generality of the following discussion, we will now discuss how to estimate weights on a group $\mathcal{U}$ such that this group is balanced with a group $\mathcal{V}$. As shown in Table 2, by balancing groups $\mathcal{U}$ and $\mathcal{V}$, it is possible to estimate different treatment effects. This estimation requires two executions of the optimization-based methods in the case of SATE and CATE, and only one execution in the case of SATT.

For example, to estimate the SATE, the treated subjects in $\mathcal{I}_{Z=1}$ are balanced with respect to all subjects in $\mathcal{I}$ to obtain the weights $w^{\text{treated}}$ over the treated subjects. Then, by balancing the control subjects in $\mathcal{I}_{Z=0}$ with respect to all subjects in $\mathcal{I}$, the weights $w^{\text{control}}$ over the control subjects are obtained. Finally, the SATE can be estimated using a (weighted) difference in means

$$\hat{\tau} = \frac{1}{\sum_{i \in \mathcal{I}_{Z=1}} w_i^{\text{treated}}} \sum_{i \in \mathcal{I}_{Z=1}} w_i^{\text{treated}} Y_i \\ - \frac{1}{\sum_{i \in \mathcal{I}_{Z=0}} w_i^{\text{control}}} \sum_{i \in \mathcal{I}_{Z=0}} w_i^{\text{control}} Y_i \quad (7)$$

or by plugging these weights in a mixed method (e.g., weight-adjusted regression similar to regression with propensity adjustment as in Rosenbaum & Rubin, 1983).

### 5.1. A seminal model

We begin our review with one of the earliest examples of an optimization-based method. The mixed integer programming for matching (MIPMatch) method (Zubizarreta, 2012) allows to match subjects with the use of an optimization model instead of the network algorithms that are traditionally used in matching methods. As a common feature of optimization-based methods, this method allows plenty of flexibility in the balance characteristics pursued without changing the optimization procedure. In contrast, the network algorithms are often designed or limited to specific matching methods.

In the general optimization model of MIPMatch, $\mathcal{U}$ and $\mathcal{V}$ are two sets of subjects as defined in Table 2, $\mathcal{O}$ and $\mathcal{C}$ are two sets of measures of covariate imbalance, $d_{ij}$ is the distance between subjects $i$ and $j$, $a_{ij}$ is a binary decision variable taking value 1 if subject $i$ is matched to subject $j$, $\omega_o$ is an importance weight, $\mu_o(\cdot)$ and $\nu_c(\cdot)$ are functions that define measures of imbalance with $a$ as the matrix of $a_{ij}$'s, $\epsilon_c$ is a scalar tolerance and $m$ is the number of subjects in $\mathcal{U}$ that needs to be matched to a subject in $\mathcal{V}$. This general optimization model is formally defined as follows:

$$\text{minimize} \quad \sum_{i \in \mathcal{U}} \sum_{j \in \mathcal{V}} d_{ij} a_{ij} + \sum_{o \in \mathcal{O}} \omega_o \mu_o(a) \tag{8}$$

subject to

$$\nu_c(a) \leq \epsilon_c, \ c \in \mathcal{C} \tag{9}$$

$$\sum_{i \in \mathcal{U}} a_{ij} = m, \ j \in \mathcal{V} \tag{10}$$

$$\sum_{j \in \mathcal{V}} a_{ij} \leq 1, \ i \in \mathcal{U} \tag{11}$$

$$a_{ij} \in \{0, 1\}, \ i \in \mathcal{U}, j \in \mathcal{V}. \tag{12}$$

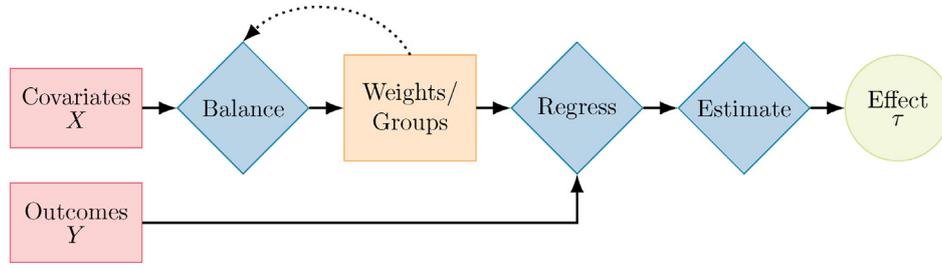

**Fig. 3.** Mixed methods.

The first term in the objective function, (8), and constraints (10)–(12) correspond to a typical matching method without replacement and with a fixed 1:$m$ ratio, and where the sum of distances is minimized. Note that constraints (10) force the fixed 1:$m$ ratio and that constraints (11) forbid matching with replacement. The second term in the objective function, (8), and constraints (9) are what allows additional flexibility.

As described in Zubizarreta (2012), several measures of covariate imbalance can be introduced in the objective function or as constraints. For example, it is possible to directly balance the univariate means by setting

$$\sum_{o \in \mathcal{O}} \omega_o \mu_o(a) = \sum_{o \in \mathcal{O}} \omega_o \left| \frac{1}{m|\mathcal{U}|} \sum_{i \in \mathcal{U}} \sum_{j \in \mathcal{V}} X_{io} a_{ij} - \frac{1}{|\mathcal{V}|} \sum_{j \in \mathcal{V}} X_{jo} \right|, \tag{13}$$

where $\mathcal{O}$ represents a subset of the observed covariates in the vector $X_i$ and $X_{io}$ corresponds to a specific covariate, and where the outer $|\cdot|$ denotes the absolute value and the inner $|\cdot|$'s again denote the cardinality of sets. While the right-hand side of (13) is not linear and hence the model cannot be solved as a mixed integer linear program in this form, it can easily be linearized by adding additional decision variables. With (13) in the objective function, the model describes a trade-off between the total sum of distances and these univariate mean's measures. Therefore, the total sum of distances unavoidably increases in exchange for an improved balance in these univariate means. Note that this trade-off can be regulated through the weights $\omega_o$'s.

Instead of putting these measures of univariate mean in the objective function, it is also possible to add them as constraints. For example, we could replace the constraints (9) with

$$\left| \frac{1}{m|\mathcal{U}|} \sum_{i \in \mathcal{U}} \sum_{j \in \mathcal{V}} X_{ic} a_{ij} - \frac{1}{|\mathcal{V}|} \sum_{j \in \mathcal{V}} X_{jc} \right| \leq \epsilon_c, \ c \in \mathcal{C}, \tag{14}$$

where $\mathcal{C}$ now also represents a subset of the observed covariates in the vector $X_i$. Note again that the constraints (14) are not linear but can be easily linearized. With (14), the model now guarantees that, in the returned solution, the mean of the covariates in $\mathcal{C}$ for the subjects in $\mathcal{U}$ are almost equal (within some tolerance) to the mean of the covariates for the subjects in $\mathcal{V}$, as long as such a balance is possible.

Other moments could also be balanced either as an objective term or as a constraint. For example, it is possible to balance the correlation of two covariates, or other measures such as quantiles or a coarse version of the Kolmogorov-Smirnov statistics, a statistics that is often used to compare two distributions (Zubizarreta, 2012). Several types of measure can be included in an optimization-based method; note however that some non-linear measures (e.g., non-convex measures) may lead to an optimization model that is difficult if not impossible to solve for even small datasets.

Furthermore, note that there exists a direct correspondence between the $a_{ij}$'s of (8)–(12) and the $w_i$'s of (7), i.e., $w_i = \sum_{j \in \mathcal{V}} a_{ij}, \forall i \in \mathcal{U}$. In other words, these weights over the subjects in $\mathcal{U}$ correspond to the number of matches in which they appear, here restricted to $w_i \in \{0, 1\}$ since the model is matching without replacement. Interpreting these matches as weights over the subjects opens the door to a variety of methods in which the allowable values of these $w_i$'s are relaxed, and different additional terms in the objective function or additional constraints are included.

In the remainder of this section, we discuss other optimization-based methods by dividing them in two categories: (1) optimized balance methods and (2) optimized attribute methods. The optimized balance methods have an objective function similar to the second term in the objective function (8) of MIPMatch, while the optimized attribute methods have constraints similar to the constraints (9) of MIPMatch. For both categories, we provide a general formulation that includes most if not all of the identified optimization-based methods.

### 5.2. Optimized balance methods

These methods obtain the balancing weights by minimizing a discrepancy measure between the weighted group $\mathcal{U}$ and the group $\mathcal{V}$, subject to constraints on the allowable weights. They can be formulated as follows:

$$\text{minimize} \quad f(w, X) \tag{15}$$

subject to

$$w \in \mathcal{W} \tag{16}$$

**Table 3**
Weight set $\mathcal{W}$ (based on Kallus, 2017a).

| Name | Definition |
|---|---|
| $\mathcal{W}^{\text{general}}$ | $\mathbb{R}^{|\mathcal{U}|}$ |
| $\mathcal{W}^{\text{simplex}}$ | $\left\{ w \in \mathbb{R}_+^{|\mathcal{U}|} : \sum_{i \in \mathcal{U}} w_i = 1 \right\}$ |
| $\mathcal{W}^{\text{b-simplex}}$ | $\mathcal{W}^{\text{simplex}} \cap [0, b]^{|\mathcal{U}|}$ |
| $\mathcal{W}^{n'\text{-subset}}$ | $\mathcal{W}^{\text{simplex}} \cap \{0, 1/n'\}^{|\mathcal{U}|}$ |
| $\mathcal{W}^{\geq \underline{n}'\text{-subset}}$ | $\cup_{n'=\underline{n}'}^{|\mathcal{U}|} \mathcal{W}^{n'\text{-subset}}$ |
| $\mathcal{W}^{n'\text{-multisubset}}$ | $\mathcal{W}^{\text{simplex}} \cap \{0, 1/n', 2/n', \ldots\}^{|\mathcal{U}|}$ |

where the objective function, (15), is a measure of discrepancy (i.e., imbalance) and the constraints, (16), specify the allowable weight vector (i.e., the values for the weights $w_i$ that can be searched to minimize the discrepancy). A list of possible weight sets for the constraints (16) is provided in Table 3. We now discuss some of these optimized balance methods.

The balance optimization subset selection (BOSS) method (Nikolaev et al., 2013; Sauppe et al., 2014; Tam Cho, Sauppe, Nikolaev, Jacobson, & Sewell, 2013) uses covariates that have been a priori discretized into bins, and selects subjects in $\mathcal{U}$ in such a way that in each bin, the number of selected subjects in $\mathcal{U}$ is as close as possible to the number of subjects in $\mathcal{V}$. Here, a covariate's bin is either a range for a continuous covariate or one to many values for a dichotomous or categorical covariate. One formulation for the BOSS method is as follows

$$\text{minimize} \quad \sum_{j \in \mathcal{P}} \sum_{k \in \mathcal{N}_j} \frac{\left[ \left( |\mathcal{V}| \sum_{i \in \mathcal{U} \cap \mathcal{B}_{jk}} w_i \right) - |\mathcal{V} \cap \mathcal{B}_{jk}| \right]^2}{\max \left\{ |\mathcal{V} \cap \mathcal{B}_{jk}|, 1 \right\}} \quad (17)$$

subject to

$$w \in \mathcal{W}^{|\mathcal{V}|\text{-subset}} \quad (18)$$

where $\mathcal{P}$ denotes the set of indices for the measured covariates, $\mathcal{N}_j$ denotes the set of indices for histogram bins for covariate $j \in \mathcal{P}$, $\mathcal{B}_{jk}$ denotes the set of subjects in the $k$th histogram bin for covariate $j$, and $\mathcal{W}^{|\mathcal{V}|\text{-subset}}$ denotes a subset of cardinality $|\mathcal{V}|$ as defined in Table 3. This formulation minimizes the standardized squared difference between the number of selected subjects in $\mathcal{U}$ and the number of subjects in $\mathcal{V}$ having a covariate value within this bin; the standardization is done by dividing by the maximum of either the number of subjects in $\mathcal{V}$ having a covariate value within this bin or 1. Several other objective functions have been proposed to measure the differences in the number of subjects over all bins. For the set of allowable weights, BOSS uses $\mathcal{W}^{|\mathcal{V}|\text{-subset}}$ (i.e., selection without replacement), as defined in Table 3, to estimate the SATT. However, it is also possible to use $\mathcal{W}^{|\mathcal{V}|\text{-multisubset}}$ (i.e., selection with replacement), which allows the estimation of the SATE; in this case, since the group $\mathcal{V}$ is larger than the group $\mathcal{U}$, it is needed to reuse the subjects in $\mathcal{U}$ multiple times. Another alternative to estimate the SATE is to sample subjects in $\mathcal{V}$ so that the new group $\mathcal{V}$ is smaller than or equal to the size of the group $\mathcal{U}$. Note that BOSS only balances the marginal distributions of the covariates.

Similarly to the BOSS method, the mutual information based matching (MIM) method (Sun & Nikolaev, 2016) selects subjects while taking into account the number of subjects per bin. However, this method now minimizes the mutual information between the covariates and the treatment variable. Furthermore, an approximate sequential selection algorithm that runs in polynomial time in the problem size is proposed for this method which improves significantly the computational time, while retaining a balancing performance similar to that of BOSS.

The kernel optimal matching (KOM) method (Kallus, 2017a; 2017b) instead minimizes the discrepancy in a reproducing kernel Hilbert space (Aronszajn, 1950). This can be done through the following formulation

$$\text{minimize} \quad \sum_{i,j \in \mathcal{U}} w_i k(X_i, X_j) w_j - \frac{2|\mathcal{U}|}{|\mathcal{V}|} \sum_{i \in \mathcal{U}} w_i \sum_{j \in \mathcal{V}} k(X_i, X_j) \quad (19)$$

subject to

$$w \in \mathcal{W}^{\text{simplex}} \quad (20)$$

where $k$ is a kernel such as the Gaussian kernel

$$k(X_i, X_j) = \exp\left( \frac{-\|X_i - X_j\|_2^2}{2\sigma^2} \right) \quad (21)$$

with bandwidth $\sigma$. Note that $\|\cdot\|_2^2$ denotes the $l_2$-norm squared. In other words, it weights the subjects in $\mathcal{U}$ so that they are balanced with respect to the subjects in $\mathcal{V}$ as measured by an objective function using kernels. This method can be used with a variety of the weight sets described in Table 3. A major advantage of using kernels such as the Gaussian kernel is that an infinite number of moments of the covariates can be balanced, in contrast to other methods that only balance a limited number of moments. In addition to the estimation of the SATT, specific formulations of KOM have been proposed for the estimation of the SATE as well as other treatment effects (Kallus, Pennicooke, & Santacatterina, 2018; Kallus & Santacatterina, 2019). These other formulations allow the estimation of the weights in one execution, instead of two as in Table 2, and lead to better results in general.

### 5.3. Optimized attribute methods

Following Wang & Zubizarreta (2019), several optimization-based methods can also be formulated to estimate balancing weights with particular properties using the following formulation:

$$\text{minimize} \quad \sum_{i \in \mathcal{U}} g(w_i) \quad (22)$$

subject to

$$\left| \sum_{i \in \mathcal{U}} w_i \phi_k(X_i) - \frac{1}{|\mathcal{V}|} \sum_{j \in \mathcal{V}} \phi_k(X_j) \right| \leq \delta_k, \ k = 1, \ldots, K \quad (23)$$

$$w \in \mathcal{W} \quad (24)$$

where $g(w_i)$ is a convex function of the weight $w_i$, $\phi_k(X_i)$ ($k = 1, \ldots, K$) are functions of the covariates $X_i$, $\delta_k$ ($k = 1, \ldots, K$) are tolerance parameters and $\mathcal{W}$ is again a set restricting the space of allowable weights (see Table 3). The model (22)–(24) differs from the model (15)–(16) since the objective function is now used to obtain weights with particular attributes (e.g., weights with maximum entropy) while the balancing property is enforced through the constraints (23). We now discuss some of these models.

The entropy balancing (EBal) method (Hainmueller, 2012) minimizes the Kullback-Leibler divergence to a set of base weights $q_i$, i.e.,

$$\text{minimize} \quad \sum_{i \in \mathcal{U}} w_i \log \frac{w_i}{q_i} \quad (25)$$

while enforcing the constraints (23) with $\delta_k = 0$ ($k = 1, \ldots, K$) and restricting the weights to the set $\mathcal{W}^{\text{simplex}}$. In the absence of a strong reason to do otherwise, the base weights generally correspond to the uniform weights $q_i = 1/|\mathcal{U}|$, which leads to the weights $w$ with maximum entropy.

The calibrating weights (CAL) method (Chan, Yam, & Zhang, 2016) method is somewhat similar to EBal by minimizing different types of distances between the weights $w_i$ and uniform weights $1/|\mathcal{U}|$, with $\delta_k = 0$ ($k = 1, \ldots, K$) and $\mathcal{W}^{\text{general}}$.

In constrast to EBal, the stable balancing weights (SBW) method (Zubizarreta, 2015) instead minimizes the variance of the weights $w_i$, i.e.,

$$\text{minimize} \quad \sum_{i \in \mathcal{U}} \left( w_i - \frac{1}{|\mathcal{U}|} \sum_{j \in \mathcal{U}} w_j \right)^2 \qquad (26)$$

while restricting them again to the simplex set $\mathcal{W}^{\text{simplex}}$. SBW also allows some slack to the moment constraints (23) by setting $\delta_k \in \mathbb{R}_+$ ($k = 1, \ldots, K$). Wang & Zubizarreta (2019) proposed a way of tuning these $\delta_k$ parameters.

The kernel balancing (KBal) method (Hazlett, 2020) replaces the constraints (23) with

$$\sum_{i \in \mathcal{U}} w_i k(X_i, X_l) = \frac{1}{|\mathcal{V}|} \sum_{j \in \mathcal{V}} k(X_j, X_l), \ \forall l \in \mathcal{I} \qquad (27)$$

where $k$ is a kernel such as the Gaussian kernel. This method allows the minimization of different objective functions such as $\sum_{i \in \mathcal{U}} w_i \log w_i$ or $-\sum_{i \in \mathcal{U}} \log w_i$ while using the weight set $\mathcal{W}^{\text{simplex}}$. This method also allows to only balance the largest eigenvalues of the kernel matrix and to relax the constraints (27) with some tolerance $\delta$. With constraints (27), the KBal method is able to balance an infinite number of moments like KOM and the kernel-based covariate functional balancing (KBCFB) method (Wong & Chan, 2018), another method using kernels.

The covariate balancing scoring rules (CBSR) method (Zhao, 2019) can also be seen as an optimized attribute method. In particular, for the estimation of the SATT, this method is similar to EBal since it maximizes the entropy with $\delta_k = 0$ for constraints (23) and with non-negative weights for constraint (24); this method does not, however, constrain the weights to sum to one like $\mathcal{W}^{\text{simplex}}$ in EBal. For the estimation of the SATE, this method instead solves the following formulation:

$$\text{minimize} \quad \sum_{i \in \mathcal{I}} (w_i - 1) \log(w_i - 1) - w_i \qquad (28)$$

subject to

$$\sum_{i \in \mathcal{I}_{Z=0}} w_i \phi_k(X_i) = \sum_{j \in \mathcal{I}_{Z=1}} w_j \phi_k(X_j), \ k = 1, \ldots, K \qquad (29)$$

$$w_i \geq 1, \ i \in \mathcal{I}. \qquad (30)$$

By doing so, it is able to compute $w^{\text{treated}}$ and $w^{\text{control}}$ in one execution.

The approximate residual balancing (ARB) method (Athey et al., 2018) is a mixed method with three steps. We show the formulation for the estimation of the SATT; a similar formulation also exists for the SATE. First, the following optimization model is solved

$$\text{minimize} \quad \eta \left\| \frac{1}{|\mathcal{I}_{Z=0}|} \sum_{i \in \mathcal{I}_{Z=0}} w_i X_i - \frac{1}{|\mathcal{I}_{Z=1}|} \sum_{j \in \mathcal{I}_{Z=1}} X_j \right\|_2^2$$
$$+ (1 - \eta) \|w\|_2^2 \qquad (31)$$

subject to

$$w \in \mathcal{W}^{\text{b-simplex}} \qquad (32)$$

where $\eta$ and $b$ are tuning parameters. This formulation is equivalent to a Lagrangian relaxation of the model (22)–(24); it could also be seen as equivalent to model (15)–(16). Then, an elastic net is fitted such that

$$\hat{\alpha} = \arg\min_\alpha \left[ \sum_{i \in \mathcal{I}_{Z=0}} (Y_i - X_i \alpha)^2 + \lambda \left\{ (1 - \gamma) \|\alpha\|_2^2 + \gamma \|\alpha\|_1 \right\} \right] \qquad (33)$$

where $\lambda$ and $\gamma$ are additional parameters. Finally, the SATT is estimated with both $w$ and $\hat{\alpha}$ as

$$\hat{\tau}_{\text{SATT}} = \frac{1}{|\mathcal{I}_{Z=1}|} \sum_{i \in \mathcal{I}_{Z=1}} Y_i$$
$$- \left\{ \frac{1}{|\mathcal{I}_{Z=1}|} \sum_{j \in \mathcal{I}_{Z=1}} X_j \hat{\alpha} + \sum_{i \in \mathcal{I}_{Z=0}} w_i (Y_i - X_i \hat{\alpha}) \right\}. \qquad (34)$$

This method is similar to the binary treatment version of augmented minimax linear (AML) estimation (Hirshberg & Wager, 2019).

The covariate balancing propensity score (CBPS) method (Fan, Imai, Liu, Ning, & Yang, 2016; Imai & Ratkovic, 2014) could also be considered as an optimization-based method, even though it is fitting a propensity score, since balance constraints are included in the optimization model. In particular, for the just-identified CBPS (or exact CBPS), the propensity score only tries to achieve the following balancing constraints when estimating the SATT:

$$\sum_{i \in \mathcal{I}_{Z=0}} \frac{\pi_\beta(X_i)}{1 - \pi_\beta(X_i)} \phi_k(X_i) = \sum_{i \in \mathcal{I}_{Z=1}} \phi_k(X_i), \ k = 1, \ldots, K \qquad (35)$$

where $\pi_\beta(X_i)$ is the parametric propensity score being fitted. When $\phi(X_i) = \frac{\partial \pi_\beta(X_i)}{\partial \beta}$, constraints (35) are equivalent to the first-order condition of the maximum log-likelihood function. In other words, when balancing the covariates that are strong predictors of treatment assignment according to the propensity score model, we recover the traditional propensity score. Note that $\frac{\pi_\beta(X_i)}{1 - \pi_\beta(X_i)}$ consists in our weight $w_i^{\text{control}}$. For the over-identified CBPS (or over CBPS), constraints (35) are then supplemented with the typical goal that the estimated propensity score should well predict the observed treatment assignment.

To conclude this subsection, we would like to highlight some notable properties of the optimization-based methods. First, some of these methods can be shown to be doubly robust; for example see Zhao & Percival (2017) in the case of EBal. This doubly robustness property is interesting since it typically leads to unbiased estimates of effects. Second, some of these optimization-based methods (e.g., ARB, AML) are used in combination with a regression on the observed outcomes (i.e., as mixed methods) instead of only using a simple weighted mean to estimate the treatment effect; these are however the exceptions as most optimization-based methods consist in balancing methods. Finally, the dual of the model (22)–(24) with $\mathcal{W}^{\text{general}}$ can be shown to be equivalent to fitting a model for the inverse propensity score with $l_1$ regularization (Wang & Zubizarreta, 2019). As a result, $\delta_k$ ($k = 1, \ldots, K$) in (23) are the dual correspondence of the $l_1$ regularization parameters; they are thus controlling the bias-variance trade-off of the estimators. Additionally, since the objective function (22) is related to the parametric form of the propensity score model, the optimization-based methods of this form are not too far in spirit to traditional methods that use the propensity score.

## 6. Comparative analysis

In this section, we provide an empirical comparison of the methods discussed above. While there exist such comparative studies for causal inference techniques (e.g., Dorie et al., 2019; Austin, 2014; Lunceford & Davidian, 2004), the optimization-based methods have not been scrutinized this way, to the best of our knowledge.

## 6.1. Data and comparative methods

It is common to use *constructed observational studies* for comparative purposes. These include studies where (1) data on the treated subjects come from a randomized controlled experiment while data on the control subjects come from an alternative data source (Cook, Shadish, & Wong, 2008; Hill, Reiter, & Zanutto, 2004; LaLonde & Maynard, 1987) and (2) an observational study is run in parallel to a randomized controlled experiment with the subjects being able to select among the two (Shadish et al., 2008). Constructed observational studies have the advantage of being representative of real data in the covariates, treatment assignment and response surface. While there exist a few of these datasets, it is unclear whether they satisfy strong ignorability.

An alternative consists of settings where the covariates come from real data, while the treatment assignment and response surface are simulated. This alternative allows to have realistic covariate distributions, while knowing the true treatment effect of interest. These datasets, however, may not allow a representative comparison of the methods in practice when an insufficient number of simulated settings are used. In this paper, we use this latter alternative. In particular, we use the 7700 datasets of the 2016 ACIC competition (Dorie et al., 2019), the largest undertaking to compare the prevailing methods.

The datasets of the 2016 ACIC competition are based on a subset of the subjects and covariates of the Collaborative Perinatal Project (Niswander & Gordon, 1972); in particular, 4802 subjects with 58 covariates, where three are categorical, five are binary, 27 are count data and the remaining 23 are continuous. Both the (binary) treatment assignment and the (continuous) response are simulated, while ensuring that strong ignorability holds in all datasets. Note that not all covariates may be confounders (i.e., need to be adjusted for). These 7700 datasets consist in 100 independent realizations of 77 different scenarios, that vary by (1) the degree of nonlinearity of the treatment assignment and response, (2) the percentage of treated subjects, (3) the overlap between the treated and control covariates, (4) the alignment between the treatment assignment and response, (5) the treatment effect heterogeneity (i.e., whether the two response curves are parallel), and (6) the magnitude of the treatment effect with respect to the noise. The reader can refer to Appendix A.1 of Dorie et al. (2019) for the details of this simulation framework. These datasets are available through a R package at https://github.com/vdorie/aciccomp/tree/master/2016.

For this empirical comparison, we consider the optimization-based methods (1) that have a public implementation available (from Comprehensive R Archive Network package repository (CRAN), GitHub, etc.), (2) that are supported by an unpublished (e.g., arXiv) or published article, and (3) that directly allow the estimation of the SATT. We restrict our focus to these methods because these are the ones that an applied researcher can readily use, while being confident that they are scientifically sound and correctly implemented. In addition, since the 2016 ACIC competition focused on the SATT, the selected optimization-based methods need to be able to estimate the SATT. With these criteria, we identify and compare the optimization-based methods in Table 4. The table depicts their R package and version, and the function used to estimate the SATT. The package for the ARB method is available at https://github.com/swager/balanceHD. The package for the KBal method is available at https://github.com/chadhazlett/KBAL. All other packages are available through CRAN. Note that, while MIPMatch respects these criteria, it is excluded from this analysis since this method with default values corresponds to a matching method and not to an optimization-based method.

Note also that the genetic matching (GenMatch) method (Diamond & Sekhon, 2013) is included, while this method is not discussed in Section 5. This method uses a genetic search algorithm to optimize the weights *over the covariates* that are used in a weighted version of Mahalanobis matching. Once satisfying weights are obtained according to a postmatching covariate balance measure, the matches are returned and the treatment effect is estimated. It is debatable whether this method is an optimization-based method since it optimizes weights that are then used in a matching method. In particular, it automates the matching with the use of optimization but we believe that it only indirectly addresses the balancing. Yet, we include this method in this comparison for comprehensiveness.

All these optimization-based methods are executed using their default parameters on R 4.0.2 to demonstrate their usefulness when directly used by an applied researcher. In addition to these default implementations, we also test the CBPS method where the method parameter is changed from "over" to "exact". Thus, in total, seven optimization-based methods are compared, later referred as `balancehd`, `cbps_over`, `cbps_exact`, `ebal`, `genmatch`, `kbal` and `sbw`. Finally, note that all methods, except the `cbps_over` and `cbps_exact` methods, require some preprocessing of the covariates. In particular, three categorical covariates (common to all 7700 datasets) are each replaced by dummy variables. Following the recommendations of the authors of these packages, we use $k-1$ dummy variables for the `ebal`, `kbal` and `sbw` methods, and $k$ dummy variables for the `balancehd` and `genmatch` methods, where $k$ indicates the number of categories. The `cbps_over` and `cbps_exact` methods were able to deal directly with these categorical covariates and thus did not need such preprocessing.

## 6.2. Empirical comparison results

This section reports the results of the empirical comparison on the 7700 datasets from the 2016 ACIC competition. In particular, the results for the selected optimization-based methods are provided in Table 5, while the results of the 24 methods from the 2016 ACIC competition are provided in Table 6. These latter results were computed from the raw results of the 2016 ACIC competition available at https://github.com/vdorie/aciccomp/tree/master/bakeoff/results_orig. Both tables report (1) the number of datasets for which it was possible to estimate the SATT, (2) the mean bias, (3) the standard deviation (SD) of the bias, and (4) the root-mean-square error (RMSE). The bias and the RMSE are both computed using the estimated and true standardized SATT, i.e., the (estimated or true) SATT divided by the standard deviation of the outcomes. Formally,

$$Bias_k = \hat{\tau}_{STD,k} - \tau_{STD,k}, \ k = 1, \ldots, 7700 \quad (36)$$

$$RMSE = \sqrt{\frac{1}{7700} \sum_{k=1}^{7700} Bias_k^2} \quad (37)$$

where $\hat{\tau}_{STD}$ and $\tau_{STD}$ denote respectively the estimated and true standardized SATT for the dataset $k$. Note that only the datasets for which it was possible to estimate the SATT are used in the computation of the bias and RMSE. In addition to these results, Table 5 depicts the mean computational time per dataset, which is not available in the raw results of the 2016 ACIC competition. Then, Table 6 shows the coverage of the 95% confidence interval of the estimated SATT, i.e., the proportion of the 7700 datasets for which the confidence interval of the estimated SATT overlaps with the true SATT. Finally, note that an overview of the covariate balance for the `kbal` method as well as two other balancing methods is provided in the Material.

**Table 4**
List of optimization-based methods.

| Method | R package | Function used |
|---|---|---|
| Approximate residual balancing (ARB) | balanceHD 1.0 | `residualBalance.ate` |
| Covariate balancing propensity score (CBPS) | CBPS 0.21 | `CBPS` |
| Entropy balancing (EBal) | ebal 0.1–6 | `ebalance` |
| Genetic matching (GenMatch) | Matching 4.9–9 | `GenMatch` |
| Kernel balancing (KBal) | kbal 0.1 | `kbal` |
| Stable balancing weights (SBW) | sbw 1.1.1 | `sbw` |

**Table 5**
Results of optimization-based methods, ordered by the RMSE and mean bias.

| Method | Number of datasets | Bias | | RMSE | Time |
|---|---|---|---|---|---|
| | | Mean | SD | | Mean (sec) |
| `kbal` | 7700 | 0.036 | 0.083 | 0.091 | 2521.3 |
| `balancehd` | 7700 | 0.041 | 0.099 | 0.107 | 2.0 |
| `sbw` | 4513 | 0.041 | 0.102 | 0.110 | 254.9 |
| `cbps_exact` | 7700 | 0.041 | 0.105 | 0.112 | 6.4 |
| `ebal` | 4513 | 0.041 | 0.110 | 0.117 | 0.2 |
| `cbps_over` | 7700 | 0.044 | 0.117 | 0.125 | 17.3 |
| `genmatch` | 7700 | 0.052 | 0.141 | 0.151 | 8282.4 |

We now discuss some of these results in more detail. First, note that the `ebal` and `sbw` methods fail to estimate the SATT on 3187 datasets, in contrast to the other four optimization-based methods. The issues with `ebal` and `sbw` are due to the collinearity of the covariates. While it would be possible to remove the collinear covariates in these datasets, this would require additional steps in which non-trivial parameters are set, and therefore we did not do it.

Second, in terms of bias and RMSE, it appears that all these methods are more or less equivalent on their estimated SATTs: the mean bias appears to be around 0.04, which is quite small; the standard deviations of the bias are higher than 0.08, which shows some variability in the biases; and the RMSEs are between 0.09 and 0.16. In terms of time, most methods can estimate the SATT within a reasonable time, i.e., less than five minutes on average. Yet, some methods are quite faster than others and only require a couple of seconds, while `kbal` and `genmatch` are outliers by taking respectively approximately 42 and 138 mins per dataset; without parallelization, it would take them around 225 and 738 days to balance the 7700 datasets. Note that taking that much time to balance a dataset may or may not be problematic depending on whether several iterations of the method are needed.

Third, when comparing our default implementations of the optimization-based methods (Table 5) to the 24 methods from the 2016 ACIC competition (Table 6), it appears that our applications of the optimization-based methods are in the worst 25% of these 24 methods in terms of bias and RMSE. For example, the method `bart` (which was also combined successfully with other methods during that competition) has a mean bias of 0.002 and a RMSE of 0.018, which is much better than our applications of the optimization-based methods. This could be due to the absence of data preprocessing or to the use of the default parameters.

Fourth, an attentive reader may have noticed that the CBPS method is also part of Table 6. In fact, it appears two times, as `cbps` and as `lasso_cbps`, and, in both cases, they are a bit better than the methods in Table 5. Two potential reasons may explain the differences in these results. First, these methods were probably implemented using non-default values for the parameters, which may have helped these methods get better estimates of the SATT. Then, as discussed in Dorie et al. (2019), these methods are used differently to estimate the SATT. In the case of `cbps`, the estimated weights were used to stratify the subjects, while in parallel a linear model was fitted to make predictions for the treated subjects. The SATT was then estimated by a weighted sum of the averages in each stratum. In the case of `lasso_cbps`, a first LASSO regression was used to select covariates to include in CBPS. Then, after obtaining the estimated weights with CBPS, these were used in another LASSO regression to estimate the SATT. Thus, in both methods, the weights were used as in a mixed method, instead of a simple weighted difference in means, which may have again helped these methods produce better estimates.

**Table 6**
Results of the 24 methods from the 2016 ACIC competition, ordered by the RMSE and mean bias.

| Method | Number of datasets | Bias | | RMSE | 95% CI coverage |
|---|---|---|---|---|---|
| | | Mean | SD | | % |
| `bart_on_pscore` | 7700 | 0.001 | 0.014 | 0.014 | 88.4 |
| `bart_tmle` | 7700 | 0.000 | 0.016 | 0.016 | 93.5 |
| `mbart_symint` | 7700 | 0.002 | 0.017 | 0.017 | 90.3 |
| `bart_mchains` | 7700 | 0.002 | 0.017 | 0.017 | 85.7 |
| `bart_xval` | 7700 | 0.002 | 0.017 | 0.017 | 81.2 |
| `bart` | 7700 | 0.002 | 0.018 | 0.018 | 81.1 |
| `sl_bart_tmle` | 7689 | 0.003 | 0.029 | 0.029 | 91.5 |
| `h2o_ensemble` | 6683 | 0.007 | 0.029 | 0.030 | 100.0 |
| `bart_iptw` | 7700 | 0.002 | 0.032 | 0.032 | 83.1 |
| `sl_tmle` | 7689 | 0.007 | 0.032 | 0.032 | 87.6 |
| `superlearner` | 7689 | 0.006 | 0.038 | 0.039 | 81.6 |
| `calcause` | 7694 | 0.003 | 0.043 | 0.043 | 81.7 |
| `tree_strat` | 7700 | 0.022 | 0.047 | 0.052 | 87.4 |
| `balanceboost` | 7700 | 0.020 | 0.050 | 0.054 | 80.5 |
| `adj_tree_strat` | 7700 | 0.027 | 0.068 | 0.074 | 60.0 |
| `lasso_cbps` | 7108 | 0.027 | 0.077 | 0.082 | 30.5 |
| `sl_tmle_joint` | 7698 | 0.010 | 0.101 | 0.102 | 58.9 |
| `cbps` | 7344 | 0.041 | 0.099 | 0.107 | 99.7 |
| `teffects_psmatch` | 7506 | 0.043 | 0.099 | 0.108 | 47.0 |
| `linear_model` | 7700 | 0.045 | 0.127 | 0.135 | 22.3 |
| `mhe_algorithm` | 7700 | 0.045 | 0.127 | 0.135 | 22.8 |
| `teffects_ra` | 7685 | 0.043 | 0.133 | 0.140 | 37.5 |
| `teffects_ipwra` | 7634 | 0.044 | 0.161 | 0.166 | 35.3 |
| `teffects_ipw` | 7665 | 0.042 | 0.298 | 0.301 | 39.0 |

Fifth, note that it is not possible to directly obtain confidence intervals of the estimated SATTs for the methods in Table 5. While some of these methods did provide some functions to estimate the confidence intervals (i.e., `balancehd`, `sbw`), these did not work due to the collinearity of the covariates. While it could be possible to obtain confidence intervals with bootstrapping for all methods, we did not pursue this avenue due to the computational resources that would be needed for some methods (e.g., `kbal`) and to the inferior results in Table 5 that did not warrant such resources. Note that such a confidence interval was a requirement of the 2016 ACIC competition, which is why the mean coverage is provided for all methods in Table 6.

Finally, we decided to also test the Bayesian additive regression trees (BART) method (Hill, 2011) with the default parameters to check whether we would be able to attain the results in the 2016 ACIC competition. We use the version 2.7 of the BART package available through CRAN. With the default parameters, we obtained the same results as those of `bart` in Table 6, with a mean computational time of less than 50 s per dataset, which is more than reasonable when compared to the optimization-based methods. Therefore, this method appears better than the selected optimization-based methods. Yet, note that BART is a regression method, and as such it does not clearly separate the "design" and "analysis" stages. In fact, all methods from Table 6, except `tree_strat`, `teffects_psmatch` and `teffects_ipw`, are either regression or mixed methods; the selected optimization-based methods are comparable to `teffects_psmatch` for the mean bias and RMSE. It is still an open research question as to why the regression-based methods (using only or in part regression) appear to perform better in general. Note, however, that optimization-based methods such as adaptive hyper-box matching (Morucci et al., 2020) and matching after learning to stretch (Parikh et al., 2021) appear competitive to BART in other experiments; these are, however, not compared here since they are not yet part of the FLAME R package at the time of writing this paper.

## 7. Research opportunities

In this section, we discuss several opportunities for the proposition of new optimization-based methods. We would like to remark that very few papers in operational research (OR) journals were identified during this review (i.e., Nikolaev et al., 2013; Sauppe et al., 2014 and Kwon et al., 2019). We believe that this is, in part, due to (1) the unawareness of causal inference methods by some members of the OR community and (2) the somewhat narrow scope of some OR journals. It is our hope that future OR work will be able to improve over the prevailing optimization-based methods for estimating causal effects, as there is room for improvement.

Clearly, the results obtained in Section 6 apply only to the seven methods, which are the publicly available implementations identified during this review. We encourage the researchers to make their implementation publicly available in the future to facilitate empirical comparison. This will also improve the impact of their work through the use of their method by applied researchers and practitioners.

### 7.1. Optimization-based mixed methods

Most of optimization-based methods discussed in Section 5 focus on the optimization of the weights without paying attention to how these weights are then used to estimate the causal effects; these methods typically use these weights with Eq. (7) to estimate a causal effect. The few exceptions are the ARB (Athey et al., 2018) and related AML (Hirshberg & Wager, 2019) mixed methods that use optimization-based weights in conjunction with a regression on the observed outcomes.

It appears reasonable to think that using such optimization-based *mixed* methods could improve over the results in Table 5, since the best methods in Table 6 are either regression or mixed methods. In particular, the best method in Table 6 is a mixed method that uses the estimated propensity score as an additional covariate in the regression model of the response. It is, however, not necessarily possible to directly replicate this method with optimization-based weights since, when estimating the SATT, these weights are all $\frac{1}{|\mathcal{I}_{Z=1}|}$ for the treated subjects and hence this additional covariate would be of no value for the regression model. Additional research is warranted to identify the best ways to use these optimization-based weights in mixed methods for the different treatment effects. In particular, replacing propensity score weights by optimization-based weights in doubly robust methods is a promising research avenue, due to the link between the propensity score weights and the optimization-based weights. Another potential avenue would be to design metaalgorithms that use these optimization-based weights similarly to the metaalgorithms by Künzel, Sekhon, Bickel, & Yu (2019) and Nie & Wager (2021) that use regression methods.

### 7.2. Other causal effects

Note that most methods identified in this review focus on the estimation of the SATT. While Table 2 does provide a framework to also estimate the SATE and the CATE with any method, it may not be the best approach to estimate these. For example, the CBSR method (Zhao, 2019) and the extensions to the KOM method proposed by Kallus et al. (2018) and Kallus & Santacatterina (2019) allow direct estimation of the SATE in one execution instead of two. We believe that such methods are preferable. In the case of the CATE, we believe that an optimization-based method that would allow the estimation of the weights without having to resolve for each value of the covariates $x$ would also be preferable. Thus, additional research in that sense is warranted, with comprehensive numerical experiments in these settings.

### 7.3. Other treatment types

Up to now, all methods focused on the typical binary treatments, i.e., $Z \in \{0, 1\}^{|\mathcal{I}|}$. Yet, several other treatment types exist such as continuous, semicontinuous, count, ordinal, multivariate and time-varying treatments.

In that respect, Fong, Hazlett, & Imai (2018) proposed the nonparametric covariate balancing generalized propensity score (npCBGPS) method, an extension of the CBPS method, for the case where the treatment is continuous, i.e., where the potential outcomes correspond to the dose-response function $Y_i(z)$, with a continuous treatment value $Z_i = z$. This method uses the following formulation

$$\text{minimize} \quad \sum_{i \in \mathcal{I}} \log w_i \qquad (38)$$

subject to

$$\sum_{i \in \mathcal{I}} w_i f(X_i^*, Z_i^*) = 0 \quad (39)$$

$$\sum_{i \in \mathcal{I}} w_i X_i^* Z_i^* = 0 \quad (40)$$

$$\sum_{i \in \mathcal{I}} w_i = |\mathcal{I}| \quad (41)$$

$$w_i \geq 0, \ \forall i \in \mathcal{I} \quad (42)$$

where $X_i^*$ and $Z_i^*$ are respectively the covariates and continuous treatment variables that have been preprocessed to have zero mean and unit variance, and $f(\cdot)$ is some function of $X_i^*$ and $Z_i^*$. This method is somewhat similar to several of the previously discussed optimization-based methods, and the resulting weights are equivalent to stabilizing inverse (generalized) propensity score weights. These weights can be used, for example, to estimate the SATE for these continuous treatment values using a (weighted) sum, i.e., $\sum_{i \in \mathcal{I}} w_i Y_i$. Note, however, that the objective function in this formulation is not convex and thus there is no guarantee that the optimization procedure will find the global optimum.

Another example of an optimization-based method that goes beyond the binary treatments is the covariate association eliminating weights (CAEW) method (Yiu & Su, 2018) which allows to estimate causal effects for semicontinuous, count, ordinal, multivariate and time-varying treatments. This method works by optimizing weights to correct a fitted propensity score function. This framework generalizes several frameworks such as the one by Fong et al. (2018).

Due to the scarcity of optimization-based methods for other treatment types and to the need for such methods, we believe that this research avenue is promising.

### 7.4. Bias-variance trade-off

The use of the tolerance parameters $\delta_k$ ($k = 1, \ldots, K$) in the constraints (23) were first introduced in the SBW method. The primary goal of these tolerance parameters is to relax the optimization such that it is feasible. In addition, these parameters can be seen as controlling the bias-variance trade-off of these estimators. In fact, Wang & Zubizarreta (2019) showed that these parameters are equivalent to regularization parameters for the model of the inverse propensity score. We believe that stochastic programming or robust optimization could be used to enhance this bias-variance trade-off. For example, we could limit the number of constraints (23) that are not perfectly balanced.

Another interesting research avenue is to derive bounds on an estimated effect instead of a single-point estimate in order to be more confident in the direction and strength of the effect (e.g., see Morucci & Rudin, 2020 in the case of matching). Similarly, developing hypothesis tests which are robust to the causal inference method used is also a promising research area (e.g., Morucci, Noor-E-Alam, & Rudin, 2021).

### 7.5. Other related settings

Another promising research area consists of other settings that are related to the current causal inference framework. These include "(1) estimating the effects of repeated interventions over time, (2) intervening on multiple dependent data instances, (3) manipulating multiple treatment variables simultaneously, (4) estimating the temporal trajectory of an outcome variable, (5) considering multiple outcome variables (perhaps with constraints or an overall cost function on their joint values), or (6) some combination of these elements" (Jensen, 2019). These also include missing data problems that are closely related to causal inference, for which some optimization-based methods (e.g., Graham, De Xavier Pinto, & Egel, 2012; Zubizarreta, 2015) have already been proposed. In addition, other interesting avenues include instrumental variables (Canan, Lesko, & Lau, 2017) and regression discontinuity (Linden & Adams, 2012) settings where the propensity score model could be replaced by weights coming from an optimization-based method (e.g., Awan et al., 2020a). Finally, note that optimization-based methods can also be used to balance groups (Bertsimas, Johnson, & Kallus, 2015; Kallus, 2018) or take into account the network interference (Awan et al., 2020b) in randomized experiments.

## 8. Conclusion

Several optimization-based methods have been proposed for estimating a treatment effect in the recent years. These appear promising since they allow to directly optimize several balance measures, eliminating or at least reducing the need for several iterations. In addition, these methods are reported to perform quite well within the computational settings they were tested. However, they were never compared extensively against each other and with other state-of-the-art methods. In this paper, we find that these methods do not perform as well as one would think. In particular, when compared against methods that model the response surface (i.e., regression or mixed methods), the optimization-based methods are inferior. They do appear, however, to perform equivalently to other balancing methods. Therefore, if an applied researcher is interested in separating the "design" and "analysis" stages, and wants to reduce the need for several iterations in the design stage, the optimization-based methods constitute an interesting option to consider. We thus propose several relevant opportunities to guide future research around these optimization-based methods.

### Acknowledgment

This research was partly funded by a grant from the Natural Sciences and Engineering Research Council of Canada (NSERC) to the second author. The lead author is also grateful for support from the Fonds de Recherche du Québec – Nature et technologies (FRQNT). This research was enabled in part by support provided by Calcul Québec (www.calculquebec.ca) and Compute Canada (www.computecanada.ca). Thanks are also due to the referees for their valuable comments.

### Supplementary material

Supplementary material associated with this article can be found, in the online version, at doi:10.1016/j.ejor.2022.01.046.